\documentclass[10pt,conference]{IEEEtran}
\usepackage{cite}
\usepackage{amsmath}
\usepackage{algorithmic}
\usepackage{algorithm}
\usepackage{array}
 \usepackage{graphicx} 
 \usepackage{booktabs}

\usepackage{url}
\begin{document}

\title{Distributed Expectation Propagation for Multi-Object Tracking over Sensor Networks}

\author{
    Qing Li$^{1,2}$, Runze Gan$^{1,2}$, James R. Hopgood$^{2}$, Michael E. Davies$^{2}$, Simon J. Godsill$^{1}$ \\ 
    $^{1}$University of Cambridge, Cambridge, U.K. \\
    $^{2}$University of Edinburgh, Edinburgh, U.K \\
     \{ql289, rg605, sjg30\}@cam.ac.uk, \{mike.davies, james.hopgood\}@ed.ac.uk
}

% \author{\IEEEauthorblockN{Qing Li, Runze Gan, Simon Godsill, James Hopgood, Michael Davies}
%  \IEEEauthorblockA{Engineering Department,\\ University of Cambridge\\
%  \{ql289, rg605, sjg30\}@cam.ac.uk}
%  }

\maketitle

\begin{abstract}
In this paper, we present a novel distributed expectation propagation algorithm for multiple sensors, multiple objects tracking in cluttered environments. The proposed framework enables each sensor to operate locally while collaboratively exchanging moment estimates with other sensors, thus eliminating the need to transmit all data to a central processing node. Specifically, we introduce a fast and parallelisable Rao-Blackwellised Gibbs sampling scheme to approximate the tilted distributions, which enhances the accuracy and efficiency of expectation propagation updates. Results demonstrate that the proposed algorithm improves both communication and inference efficiency for multi-object tracking tasks with dynamic sensor connectivity and varying clutter levels.
%This paper presents a novel distributed expectation propagation algorithm for multi-sensor, multi-object tracking in cluttered environments. The proposed distributed expectation propagation framework enables each sensor to operate locally while collaboratively exchanging moment estimates with other sensors, eliminating the need to transmit all data to a central processing node. Specifically, we introduce a fast Rao-Blackwellised Gibbs sampling scheme to approximate the tilted distributions to enhance the accuracy and efficiency of expectation propagation updates.  Results demonstrate the proposed expectation propagation algorithm’s advantage to enhance communication and inference efficiency for multi-object tracking tasks with dynamic sensor connectivity and varying clutter levels.
\end{abstract}

% no keywords
%\IEEEpeerreviewmaketitle

\section{Introduction}
Multi-sensor multi-object tracking (MS-MOT) is a fundamental task  in applications such as autonomous systems and surveillance. Traditional MS-MOT approaches \cite{bar1995multi,koch2016tracking} focus on centralised fusion strategies, where sensors transmit all data to a central processing unit for inference tasks. However, these methods face significant challenges in communication constraints, and practically sensors often can only access local data due to privacy concerns.
Therefore, various distributed tracking algorithms have been proposed \cite{rao1993fully,olfati2005distributed,chong1990distributed, sandell2008distributed,chong2017forty}. Two suboptimal fusion rules, Generalised Covariance Intersection (GCI) \cite{clark2010robust} and Arithmetic Average (AA) \cite{li2017generalized}, can be integrated into existing multi-object trackers \cite{li2017efficient,ueney2013distributed,li2023scalable,li2020arithmetic,li2018computationally} for fusing local multi-object distributions with unknown correlations. However, these suboptimal approaches often suffer from degraded accuracy.

Recently, variational trackers \cite{gan2024variational,gan2022variational} have demonstrated their advantages in tracking accuracy and computation speed, and have been explored for distributed tracking from an optimal fusion perspective. Specifically, a fully decentralised counterpart of the centralised multi-sensor variational tracker was developed in \cite{li2023consensus}. %, which is mathematically equivalent to the centralised fusion when consensus arrives. 
Subsequently, a decentralised gradient-based variational inference framework \cite{li2024decentralised} has shown to further lower the communication cost in comparison to \cite{li2023consensus}, which also allows sensors to operate independently without requiring consensus during variational inference iterations.

%In this paper, we focus on improving communication efficiency of the distributed multi-object tracker by minimising inter-sensor message-passing iterations. Specifically, we use the expectation propagation (EP) framework \cite{vehtari2020expectation,xu2014distributed} to design a distributed implementation of the fast Rao-Blackwellised Gibbs sampling tracker in \cite{li2023adaptive,li2022scalable}. It is ,to our knowledge, the first paper to to design a fully distributed EP with no central node for MS-MOT tasks under the Non-homogeneous Poisson Process (NHPP) measurement model. 

In this paper, we address the challenge of improving communication efficiency in distributed multi-object tracking by minimising inter-sensor message-passing iterations. To achieve this, we develop a distributed implementation of the fast Rao-Blackwellised Gibbs sampling tracker \cite{li2023adaptive,li2022scalable} using the expectation propagation (EP) framework \cite{vehtari2020expectation,xu2014distributed}. To our knowledge, this is the first work to design a  distributed EP approach for multi-sensor multi-object tracking (MS-MOT) under the Non-homogeneous Poisson Process (NHPP) measurement model. The proposed EP framework supports both centralised and fully decentralised implementations, with the latter relying solely on local communication between neighbouring nodes. %, making it well-suited for scalable systems.
Results show that our approach maintains high tracking accuracy with fewer than ten EP iterations in our experiments,  %It also shows that compared to variational inference methods, the proposed distributed EP algorithms typically converges in fewer iterations, 
making it well-suited for scenarios where minimising inter-sensor communication is a priority.

\section{Model and assumptions}\label{sec:model}

\subsection{Dynamical  Model}\label{dynamic model}
Consider that there are $K$ targets in the surveillance area and $K$ is known. At time step $n$, their joint state is $X_n=[X_{n,1}^\top,...,X_{n,K}^\top]^\top$. Assume that targets move in a 2D area,  i.e.,  $X_{n,k}=[x^1_{n,k}, \Dot{x}^1_{n,k},x^2_{n,k}, \Dot{x}^2_{n,k}]^\top$, $k\in \{1,...,K\}$, where $x^d_{n,k}$ and $\Dot{x}^d_{n,k}$, $d\in\{1,2\}$ are the $k$-th target's position and velocity in the $d$-th dimension, respectively. We assume an independent linear Gaussian transition density:% for each target's states:
\vspace{-0.4em}
\begin{equation}\label{eq: dynamic transition}
    p(X_n|X_{n-1})
    =\prod\nolimits_{k=1}^K\mathcal{N}(X_{n,k};F_{n,k}X_{n-1,k},Q_{n,k}). %\\[-0.4em]
\end{equation}
where $F_{n,k}=\text{diag}(F^1_{n,k},F^2_{n,k})$, $Q_{n,k}=\text{diag}(Q^1_{n,k},Q^2_{n,k})$. 
%For a constant velocity (CV) model, $F_{n,k}^d,Q_{n,k}^d$ ($d=1,2$) are
% \vspace{-0.5em}
% \begin{equation} \label{eq:model para}
%     F_{n,k}^d=\begin{bmatrix} 1 & \tau \\ 0& 1
%     \end{bmatrix}, 
%     Q_{n,k}^d=\sigma_k^2\begin{bmatrix} \tau^3/3 & \tau^2/2 \\ \tau^2/2& \tau
%     \end{bmatrix}, 
% \end{equation}
% where $\tau$ is the time interval between time steps.

\subsection{Measurement Model and Association Prior}\label{NHPP measurement model and association prior}
We consider $N_s$ sensors monitoring the surveillance area, each with an NHPP measurement model as in \cite{gilholm2005poisson}. The parameters of the NHPP model may differ across sensors.
Denote the measurements from all sensors at time step $n$ be $Y_n=[Y_{n}^{1},...,Y_{n}^{N_s}]$. Each $Y_{n}^{s}$ includes measurements from $s$-th sensor, and $Y_{n}^{s}=[Y_{n,1}^{s},...,Y_{n,M_n^{s}}^{s}]$, where each $Y_{n,j}^{s}$, $j=1,...M_n^{s}$ is a vector representing the 2D Cartesian coordinates, and $M_n^{s}$ is the total measurement number of the $s$-th sensor. Subsequently, $M_n=[M_n^1,...,M_n^{N_s}]$ records the total number of measurements received from all sensors at time step $n$.
Define the set of Poisson rates for all sensors as $\Lambda=[\Lambda^{1},...,\Lambda^{N_s}]$. For each sensor $s$, the Poisson rate vector is $\Lambda^{s}=[\Lambda_0^{s},...,\Lambda_K^{s}]$, where $\Lambda_0^{s}$ is the clutter rate and $\Lambda_k^{s}$ is the $k$-th object rate, $k=1,...,K$.  
%For each sensor $s$, each target $k$ generates measurements by a NHPP with a Poisson rate $\Lambda_k^{s}$, and the total measurement process is also a NHPP from the superposition of the conditional independent NHPP measurement process from $K$ targets and clutter.
The total measurement number from the $s$-th sensor follows a Poisson distribution with a rate of $\Lambda_{sum}^{s}=\sum_{k=0}^K\Lambda_k^{s}$. 

Our independent measurement model assumption implies that given $X_{n}$, the measurements of each sensor are conditionally independent: $p(Y_n|X_n)=\prod_{s=1}^{N_s} p(Y_{n}^{s}|X_{n})$.
% \begin{equation}
%     p(Y_n|X_n)=\prod_{s=1}^{N_s} p(Y_{n}^{s}|X_{n})
% \end{equation}
% , in other words, the joint likelihood function can be factorised as the product of all local likelihood functions
% Define the local likelihood function at each sensor $s$ as $p(Y_{n}^{s}|X_{n})$. The joint likelihood function is the product of all local likelihood functions since all $Y_{n}^{s}$ are assumed conditionally independent given $X_{n}$, that is, $p(Y_n|X_n)=\prod_{s=1}^{N_s} p(Y_{n}^{s}|X_{n})$.
% Given $Y_n$, the associations at time step $n$ are denoted by $\theta_{n}=[\theta_{n}^{1},\theta_{n}^{2},...,\theta_{n}^{N_s}]$
We denote the association variables of all measurements $Y_n$ by $\theta_{n}=[\theta_{n}^{1},...,\theta_{n}^{N_s}]$, with each $\theta_{n}^{s}=[\theta_{n,1}^{s},...,\theta_{n,M_n^{s}}^{s}]$ ($s=1,...,N_s$) representing the association vector for the $s$-th sensor's measurements. Each component $\theta_{n,j}^{s}$ ($j=1,...,M_n^{s}$) gives the origin of the measurement $Y_{n,j}^{s}$; $\theta_{n,j}^{s}=0$ indicates that $Y_{n,j}^{s}$ is generated by clutter, and $\theta_{n,j}^{s}=k$ ($k=1,...,K$) means that $Y_{n,j}^{s}$ is generated from the target $k$. The adopted conditionally independent NHPP measurement model leads to the following properties. First, $p(Y_n,\theta_n|X_n,M_n)=p(Y_n|\theta_n,X_n)p(\theta_n|M_n)$. Both joint association prior and joint likelihood are conditionally independent across sensors, and measurements are conditionally independent given associations and states, i.e., $p(\theta_n|M_n) =\prod\nolimits_{s=1}^{N_s}  p(\theta_{n}^{s}|M_n^{s})$, $p(Y_n|\theta_n,X_n)=\prod\nolimits_{s=1}^{N_s} p(Y_{n}^{s}|\theta_{n}^{s},X_{n})$, and $p(Y_{n}^{s}|\theta_{n}^{s},X_{n})=\prod\nolimits_{j=1}^{M_n^{s}}\ell^s(Y_{n,j}^{s}|X_{n,\theta_{n,j}^{s}})$,
% %\vspace{-0.5em}
% \begin{align}\label{eq: joint association prior conditionally independent}
%    & p(\theta_n|M_n) =\prod\nolimits_{s=1}^{N_s}  p(\theta_{n}^{s}|M_n^{s})\\[-0.2em]\label{eq: joint likelihood conditionally independent}
%   &  p(Y_n|\theta_n,X_n)=\prod\nolimits_{s=1}^{N_s} p(Y_{n}^{s}|\theta_{n}^{s},X_{n})\\[-.1em]
%     \label{eq:obs prior}
% &p(Y_{n}^{s}|\theta_{n}^{s},X_{n})=\prod\nolimits_{j=1}^{M_n^{s}}\ell^s(Y_{n,j}^{s}|X_{n,\theta_{n,j}^{s}})%\\[-2em]\notag
% \end{align}
where $M_n^{s}$ is implicitly known from $\theta_n^{s}$ since $M_n^{s}$ is the cardinality of $\theta_n^{s}$, and $\ell^s$ is the probability density function of a single measurement received in sensor $s$ given its originator's state. Here we assume a linear and Gaussian model for object originated measurements and clutter measurements to be uniformly distributed in the observation area of volume $V^{s}$:
%\vspace{-0.4em}
\begin{equation} \ell^s(Y_{n,j}^{s}|X_{n,k})=\begin{cases} 
    \mathcal{N}(H X_{n,k},R_{k}^{s}),& \text{$k\neq 0 $; \ \ (object)}\\
     {1}/{V^{s}}, & \text{$k= 0 $; \ \ (clutter)}
\end{cases} 
%\\[-0.5em]
\label{measurement model}
\end{equation}
where observation matrix $H = \mathrm{diag}([1 \; 0], [1\;0])$. $R_{k}^{s}$ indicates the $s$-th sensor noise covariance. Moreover, the joint prior $p(\theta_{n}^{s}|M_n^{s})$ can be factorised as the product of $M_n^{s}$ independent association priors, i.e., $p(\theta_{n}^{s}|M_n^{s})=\prod_{j=1}^{M_n^{s}} p(\theta_{n,j}^{s})$,
where $p(\theta_{n,j}^{s})$ is a categorical distribution with $\theta_{n,j}^{s} \in \{0,...,K\}$
%\vspace{-0.3em}
\begin{align}
    \label{eq:single assoc prior}    &p(\theta_{n,j}^{s})=\frac{1}{\sum_{k=0}^K\Lambda_k^{s}}\sum\nolimits_{k=0}^K\Lambda_k^{s}\delta[\theta_{n,j}^{s}=k].
\end{align}

\begin{algorithm}
\caption{Procedure of Distributed EP for tracking}
\label{alg:EP}
\begin{algorithmic}[1]
\STATE \textbf{Initialisation}: Set $g_s(X_n) = 1$ for all sites $s$.
    \FOR{EP iteration i=1,...$I_{max}$}
\FOR{each site $s = 1, \dots, N_s$}
    \STATE \textbf{Step 1: Compute locally}:\\ 1) the cavity distribution $g_{-s}(X_n)$ by \eqref{eq:cavity distribution}\\
    2) the tilted distribution $\tilde{g}^s(X_n)$ by \eqref{eq: tilt}\\
    3) Set the moments of $g^{\text{new}}(X_n)$ to $ \tilde{g}^s(X_n)$ \\
    4) Update the site approximation $g_s(X_n)$ by \eqref{eq:updated site approximation}
    \STATE \textbf{Step 2: Pass parameters of $g_s(X_n)$
    to other sensors}
\ENDFOR
   \ENDFOR
\STATE Return approximation $ g(X_n)\propto \hat{p}_n(X_n) \prod_{s=1}^{N_s} g_s(X_n)$
\end{algorithmic}
\end{algorithm}

\section{Distributed Expectation Propagation for multi-sensor
Multi-Object Tracking}
%In this section, we detail the use of Expectation Propagation (EP) for distributed inference in a multi-sensor, multi-object tracking scenario. EP efficiently approximates the posterior distribution of shared target states while marginalising out local sensor-specific data associations. This framework allows efficient partitioning of data, enabling each sensor to operate independently, thereby reducing computational complexity and improving scalability in distributed networks.

\subsection{Problem formulation}\label{sec:Problem formulation}
%With the existence of measurement origin uncertainty,  a
A Bayesian object tracker aims recursively to estimate the posterior $p(X_n,\theta_n|Y_{1:n})$ in the 
prediction and update steps:
\vspace{-0.5em}
\begin{align} \label{eq:prediction}
    &\hspace{-0.5em} p(X_n|{Y}_{1:n-1})= \int p(X_n|X_{n-1}) p({X}_{n-1}|{Y}_{1:n-1})d{X}_{n-1}\\[-0.5em]\notag 
    &\hspace{-0.5em} p(X_n|{Y}_{1:n})\propto
    p(X_n|{Y}_{1:n-1}) \int p(Y_n|\theta_n,X_n) p(\theta_n|M_n) d\theta_n \\[-2.2em]\notag 
\end{align}
Since this exact filtering recursion is intractable due to the association uncertainty, here the posterior $p(X_n|{Y}_{1:n})$  is approximated with  $\hat{p}_n(X_{n}|Y_n)$ using EP at each time step $n$. This approximation $\hat{p}_n(X_{n}|Y_n)$ will be used to formulate the predictive prior in \eqref{eq:prediction} for the next time step.  Under model assumptions in Section \ref{sec:model}, the target posterior of EP is:
\vspace{-0.5em}
\begin{equation}\label{eq:posterior ep}
\hat{p}_n(X_{n}|Y_n)\propto \hat{p}_n(X_{n}) \prod_{s=1}^{N_s} \int p(Y_n^{s} \mid \theta_n^{s}, X_n) p(\theta_n^{s}) \, d\theta_n^{s},
\end{equation}
where $\hat{p}_n(X_{n})$ is the approximated predictive prior formulated by using the approximation distribution $g(X_{n-1})$ of $\hat{p}_{n-1}({X}_{n-1}|{Y}_{1:n-1})$ at previous time step $n-1$:
\vspace{-0.5em}
\begin{equation}  \label{eq: predictive prior}
     \hat{p}_n(X_n) = \int p(X_n|X_{n-1})g(X_{n-1})dX_{n-1}. 
\end{equation}
Here, for simplicity, the time-step subscript of $g(\cdot)$ is omitted.%, though note that $g(\cdot)$ varies with $n$.

Assume an independent initial Gaussian prior $p(X_{0})=\prod_{k=1}^K p(X_{0, k}) =\prod_{k=1}^K\mathcal{N}(X_{0,k};\mu^{k}_{0|0},\Sigma^{k}_{0|0})$. Then, given the %independent linear Gaussian
transition in \eqref{eq: dynamic transition},
its predictive prior $\hat{p}_n(X_n)$ can always maintain an independent Gaussian form, i.e., $\hat{p}_n(X_n)=\prod_{k=1}^K \hat{p}_n(X_{n,k})$, where for each object $k$, we have
%\vspace{-1.5em}
\begin{align}\label{eq:predictive prior computation}  \hat{p}_n(X_{n,k})=&\mathcal{N}(X_{n,k};\mu^{k}_{n|n-1},\Sigma^{k}_{n|n-1}),
    \\\notag
    \mu^{k}_{n|n-1}=&F_{n,k}\mu^{k}_{n-1|n-1},\\\notag
    \Sigma^{k}_{n|n-1}=&F_{n,k}\Sigma^{k}_{n-1|n-1}F_{n,k}^\top+Q_{n,k}.
\end{align}
% In the following sections, we will elaborate on how EP is used to approximate $\hat{p}_n(X_{n}|Y_n)$ in \eqref{eq:posterior ep} by iteratively refining a set of site-specific approximations.

\subsection{Expectation Propagation Framework for Tracking}
EP approximates the posterior $\hat{p}_n(X_{n}|Y_n)$  in \eqref{eq:posterior ep} with a global approximation distribution $g(X_n)$ by factorising it into a product of local factors \cite{xu2014distributed}:
\vspace{-0.5em}
\begin{equation}\label{eq:ep}
    \hat{p}_n(X_{n}|Y_n) \approx g(X_n) \propto \hat{p}_n(X_n) \prod_{s=1}^{N_s} g_s(X_n), \\[-0.5em]
\end{equation}
where $g_s(X_n)$ represents the site approximation from sensor $s$, $s=1,..., N_s$. Note that the prior $\hat{p}_n(X_n)$ is already included in the global approximation $g(X_n)$ as constructed in \eqref{eq:ep}, thus we do not need an additional site approximation $g_0(X_n)$.
% Since $\hat{p}_n(X_n)$ and $g(X_n)$ are both Gaussian, their product remains Gaussian, and we can exactly incorporate the effect of the prior into the approximate posterior without needing an additional site term $g_0(X_n)$.

For each site $s$, EP defines a \textit{cavity distribution}, which represents the current approximation without the contribution from site $s$ (indicated with subscript $-s$):
\begin{equation}\label{eq:cavity distribution}
    g_{-s}(X_n) \propto \frac{g(X_n)}{g_s(X_n)} 
\end{equation}

By using  $ g_{-s}(X_n)$, the objective is to find the site approximation $g_s(X_n)$ that minimises 
%such that  $g_s(X_n)g_{-s}(X_n)$ approximates $p(Y_n^{s} \mid X_n)g_{-s}(X_n)$, which means that 
the  Kullback-Leibler divergence $\text{KL}(p(Y_n^{s} | X_n)g_{-s}(X_n) || g_s(X_n)g_{-s}(X_n))$. To do it, we compute a tilted distribution $\tilde{g}^s(X_n)$ that incorporates the exact likelihood and the cavity distribution of $s$-th site :
\begin{align}\notag
  \tilde{g}^s(X_n) 
  &\propto g_{-s}(X_n)  p(Y_n^{s} \mid X_n) \\
  \label{eq: tilt}
   &\propto g_{-s}(X_n) \int p(Y_n^{s} \mid \theta_n^{s}, X_n) p(\theta_n^{s}) \, d\theta_n^{s}.
\end{align}
With moment matching, we construct a new approximate posterior $g^{new}(X_n)$ by setting the moments of $g^{new}(X_n)$ equal to the  tilted distribution 
 $\tilde{g}^s(X_n)$. Subsequently, the updated site approximation can be computed by:
\begin{equation}\label{eq:updated site approximation}
       g_s(X_n) \propto \frac{g^{new}(X_n)}{g_{-s}(X_n)}. 
\end{equation}
Finally, the full procedure of EP can be seen in Algorithm \ref{alg:EP}.
%We can see that the idea of EP is to minimise each local KL-divergence $\text{KL}(p(Y_n^{s} | X_n)g_{-s}(X_n) || g_s(X_n)g_{-s}(X_n))$, which means it will not in general minimize the KL-divergence from the target posterior to the global approximation $\text{KL}(\hat{p}_n(X_{n}|Y_n) || g(X_n)$.

%In addition,  when the tilted distribution $\tilde{g}_s(X_n)$ is intractable, we can use Monte Carlo sampling to obtain the mean and covariance of $\tilde{g}_s(X_n)$. Accordingly, the parameters of $g^{new}(X_n)$ is set to match these computed moments, thereby updating the site-specific approximation  $g_s(X_n)$. We will discuss a fast Rao-Blackwellised Gibbs sampling scheme in the following section.

\subsection{A Rao-Blackwellised Gibbs sampler for approximating the tilted distribution}\label{sec:Rao-Blackwellised Gibbs}
Since the tilted distribution $\tilde{g}^s(X_n)$ is intractable, this section designs an efficient Rao-Blackwellised Gibbs sampler to approximate $\tilde{g}^s(X_n)$. %, thereby obtaining the mean and covariance of $\tilde{g}_s(X_n)$ for the moment matching to the parameters of $g^{new}(X_n)$. 
%In this section, we detail how to approximate the tilted distribution $\tilde{g}^s(X_n)$ using the MCMC sampling. Specifically, we adopt an efficient Rao-Blackwellised Gibbs sampling scheme that includes an auxiliary sampling step for $\theta^s_{n}$. Therefore, instead targeting  $\tilde{g}^s(X_n)$, we choose the joint distribution $\tilde{g}^s(X_n, \theta^s_{n})$ as the stationary distribution, while only keeping the converged sample set of $X_n$ to approximate the tilted distribution $\tilde{g}^s(X_n)$.
Specifically, to enable a parallel processing and more efficient inference, instead of targeting  $\tilde{g}^s(X_n)$, we choose the joint distribution $\tilde{g}^s(X_n, \theta^s_{n})$ as the stationary distribution, while only keeping the converged sample set of $X_n$ to approximate the tilted distribution $\tilde{g}^s(X_n)$. 
% The joint distribution is written as
% \begin{equation}
%      \tilde{g}^s(X_n, \theta^s_{n}) = g_{-s}(X_n) p(Y_n^{s} \mid \theta_n^{s}, X_n) p(\theta_n^{s})
% \end{equation}
In this way, the full conditionals of $\theta^s_{n}$ and $X_{n}$ for the Gibbs sampling steps are all in closed form, and each association $\theta_{n,j}^s$, $j=1,...,M^s_n$ can be sampled in parallel with its conditional being
% The conditional distribution for association $\tilde{g}^s( \theta^s_{n}|X_n)$ is
% \begin{equation}
%      \tilde{g}^s(\theta^s_{n}|X_n) \propto p(Y_n^{s} \mid \theta_n^{s}, X_n) p(\theta_n^{s})
% \end{equation}
%Specifically, the conditional of each association $\theta_{n,j}^s$, $j=1,...,M^s_n$ can be sampled in parallel as follows %from the conditional $\tilde{g}^s( \theta^s_{n,j}|X_n)$, 
\vspace{-0.5em}
\begin{align}%\notag
 \tilde{g}^s( \theta^s_{n,j}|X_n) 
 %&= p(\theta^s_{n,j}|Y_{n,j},{X}_{n})  \\
   &= \Big(\frac{\Lambda_{0}}{V \Tilde{l}}\Big)^{[\theta_{n,j}=0]} \prod_{k=1}^{K} \Big(\frac{\Lambda_{k} l_{kj}}{\Tilde{l}}\Big)^{[\theta_{n,j}=k]} .\\[-2.2em]\notag
\end{align} 
where $l_{kj}=\mathcal{N}(Y_{n,j}^{s};H X_{n,k},R_{k}^{s})$, and $\Tilde{l}$ is a normalisation constant. $[\cdot]$ is the Iverson bracket and $[\theta_{n,j}=k]$ evaluates to 1 if $\theta_{n,j}=k$, and 0 otherwise.
Likewise, each target state $X_{n,k}$, $k=1,...,K$ can also be sampled in parallel:
\begin{align} \notag
         \tilde{g}^s(X_{n,k} | \theta^s_{n}) &
     %\propto p({Y}_{n}|{X}_{n},\theta^s_{n}) g_{-s}(X_n)  \\\notag
    %& \propto \prod_{j\in \Theta^i_n} p({Z}_{n,j}|{X}_{n,i})g_{-s}(X_{n,i}) \\\notag
     %&
     \propto \mathcal{N}(\Tilde{Z}^{k}_{n};H X_{n,k},\Tilde{R_{k}})\mathcal{N}(X_{n,k};\mu^{-s}_{k} ,\sigma^{-s}_{k})\\   \label{Sample state x-n}
    & \propto  \mathcal{N}(X_{n,k};\tilde{\mu_{k}},\tilde{\Sigma_{k}}).
\end{align}
where $ {\Tilde{Z}_{n}^k}=\frac{1}{|\Theta^k_n|}\sum_{j\in \Theta^k_n} {Z}_{n,j}$, $
    \Tilde{R_{k}}= \frac{1}{|\Theta^k_n|}R_{k} $,
    $\Theta^k_n=\{j|j\in\{1,...,M_n\},\theta_{n,j}=k\}$, and $|\cdot|$ denotes the cardinality of the set. See \cite{li2024decentralised} for detailed derivations. Finally, $\tilde{\mu_{k}},\tilde{\Sigma_{k}}$ can be computed by Kalman update equations.

Upon the Gibbs sampler converges,
%to the stationary distribution $ \tilde{g}^s(X_n, \theta^s_{n})$, the output samples $\{ \theta_{n}^{s,(p)}\}_{p=1}^{N_p}$ are implicitly drawn exactly from $\tilde{g}^s( \theta^s_{n})$. Therefore, 
we can compute the tilted distribution $\tilde{g}^s(X_n)$ by marginalising out $\theta^s_{n}$ using samples $\{ \theta_{n}^{s,(p)}\}_{p=1}^{N_p}$:
\vspace{-.5em}
\begin{equation} \label{eq:tilted approximation}
\tilde{g}^s(X_n) = \int \tilde{g}^s(X_n, \theta^s_{n}) d \theta^s_{n} \approx \frac{1}{N_p} \sum_{p=1}^{N_p} \tilde{g}^s(X_{n} | \theta^{s,(p)}_{n})  \\[-0.5em]
\end{equation}
Thus the MCMC approximation of $\tilde{g}^s(X_n)$ takes the form of a Gaussian mixture.

\subsection{Implementation of the Fully Distributed  EP}
At the previous time step $n-1$, the posterior $ \hat{p}_{n-1}(X_{n-1}|Y_{n-1})$  is approximated by $ g(X_{n-1})= \prod_{k=1}^K g(X_{n-1,k})=\prod_{k=1}^K\mathcal{N}(X_{n-1,k};\mu^{k}_{n-1|n-1},\Sigma^{k}_{n-1|n-1})$, under settings in Section \ref{sec:Problem formulation},  . 

At time step $n$,  we perform the prediction step using \eqref{eq: predictive prior}, and the predictive prior  has an independent Gaussian form of $\hat{p}_n(X_n)=\prod_{k=1}^K \hat{p}_n(X_{n,k})=\prod_{k=1}^K\mathcal{N}(X_{n,k};\mu^{k}_{n|n-1},\Sigma^{k}_{n|n-1})$, where $\mu^{k}_{n|n-1}$ and $\Sigma^{k}_{n|n-1}$ are computed by \eqref{eq:predictive prior computation}. 

Then, we perform the update step in \eqref{eq:posterior ep}. By using EP method, the posterior $\hat{p}_n(X_{n}|Y_n)$ can be approximated by a multivariate Gaussian distribution $ g(X_n)=\prod_{k=1}^{K} g(X_{n,k})=\prod_{k=1}^{K}\mathcal{N}(X_{n,k};\mu^{k}_{n|n},\Sigma^{k}_{n|n})$.
$g(X_n)$ is the global approximation; the site approximation is $g_s(X_n)$ defined in \eqref{eq:ep}, and $g(X_n) \propto \hat{p}_n(X_n) \prod_{s=1}^{N_s} g_s(X_n)$.

Here we rewrite global approximation $g(X_n)$, the prior $\hat{p}_n(X_n)$ and the site approximation $g_s(X_n)$, $s=1,..., N_s$ in the form of a canonical exponential family distribution. To start, the  global approximation
$ g(X_n; \lambda_1,\lambda_2)=\prod_{k=1}^{K} g(X_{n,k}; \lambda_{1,k},\lambda_{2,k})$,
where $\lambda_{1,k}$ and $\lambda_{2,k}$ are natural parameters and can be computed from its moment form, $\lambda_{1,k}={(\Sigma^{k}_{n|n})}^{-1}\mu^{k}_{n|n}$ and $\lambda_{2,k}={(\Sigma^{k}_{n|n})}^{-1}$ and vice-versa,  $\mu^{k}_{n|n}={(\lambda_{2,k})}^{-1}\lambda_{1,k}$, $\Sigma^{k}_{n|n}={(\lambda_{2,k})}^{-1}$.
% \begin{equation}
%     \lambda_{1,k} = {(\Sigma^{k}_{n|n})}^{-1} \mu^{k}_{n|n}, \quad \lambda_{2,k} = -\frac{1}{2} {(\Sigma^{k}_{n|n})}^{-1}.
% \end{equation}
% and vice-versa,  
% %Conversely, the moment parameters can be obtained from the natural parameters as:
% \begin{equation}
%     \mu^{k}_{n|n} = -\frac{1}{2} {(\lambda_{2,k})}^{-1} \lambda_{1,k}, \quad \Sigma^{k}_{n|n} = -\frac{1}{2} {(\lambda_{2,k})}^{-1}.
% \end{equation}
Similarly, the exponential family distribution forms of $\hat{p}_n(X_n)$ and $g_s(X_n)$ are
$\hat{p}_n(X_n;\eta_1, \eta_2)=\prod_{k=1}^{K}\hat{p}_n(X_{n,k};\eta_{1,k}, \eta_{2,k})$, $g_s(X_n; \lambda_1^s,\lambda_2^s)=\prod_{k=1}^{K} g(X_{n,k}; \lambda_{1,k}^s,\lambda_{2,k}^s)$, respectively. %, and their natural parameters and can be computed from their moment forms. 
Subsequently, the global approximation parameters $\lambda_{1},\lambda_{2}$ can be obtained by summing up all the local approximation parameters and the prior parameters\cite{xu2014distributed,williams2006gaussian}:
\begin{equation}\label{eq:global apprx}
    \lambda_{1}=\eta_1+ \sum_{s=1}^{N_s} \lambda_1^s, \ \lambda_{2}=\eta_2+ \sum_{s=1}^{N_s} \lambda_2^s \\[-0.5em]
\end{equation}
%This follows from the Gaussian multiplication rules,
%, which results in an unnormalised Gaussian distribution with updated natural parameters, 
%see \cite{williams2006gaussian}. %The exponential family form enables simple computation, as products and divisions remain within the same parametric family via summation and subtraction of natural parameters.

% Specifically, given two Gaussian densities,
% \begin{equation}
%     g_1(\theta \mid r_1, Q_1) g_2(\theta \mid r_2, Q_2) \propto g_{1 \cdot 2}(\theta \mid r_1 + r_2, Q_1 + Q_2),
% \end{equation}
% and analogously, for division:
% \begin{equation}
%     \frac{g_1(\theta \mid r_1, Q_1)}{g_2(\theta \mid r_2, Q_2)} \propto g_{1/2}(\theta \mid r_1 - r_2, Q_1 - Q_2).
% \end{equation}

Next, we introduce the iterative procedure of the EP algorithm for approximating the posterior $\hat{p}_n(X_n|Y_n)$ by $g(X_n)$.

At the initial $0$-th iteration, all site approximations $g_s(X_n)$, $s = 1, \dots, N_s$, are set with $\lambda_{1,k}^s(0) = 0$ and $\lambda_{2,k}^s(0) = 0$ for $k = 1, \dots, K$. Thus, the global approximation $g(X_n; \lambda_1(0), \lambda_2(0))$ is initialized to the prior $\hat{p}_n(X_n)$, i.e., $\lambda_1(0) = \eta_1, \quad \lambda_2(0) = \eta_2$. For each iteration $i$, the EP algorithm update each local approximation $g_s(X_n; \lambda_1^s(i),\lambda_2^s(i))$ according to Algorithm \ref{alg:EP} until it converges, although the convergence is not guaranteed. Should it converge, the local posteriors across all nodes will be the same.

Specifically, at $i$-th iteration,  the natural parameters of the cavity distribution $g_{-s}(X_n; \lambda_1^{-s}(i),\lambda_2^{-s}(i))$ are updated as
%Now we take a look at the specific steps at $i$-th iteration. According to the EP steps, for each site $s = 1, \dots, N_s$, we compute the cavity distribution $g_{-s}(X_n; \lambda_1^{-s}(i),\lambda_2^{-s}(i))$,
% \begin{equation}
%    g_{-s}(X_n; \lambda_1^{-s}(i),\lambda_2^{-s}(i)) \propto \frac{g(X_n; \lambda_1(i-1),\lambda_2(i-1))}{g_s(X_n; \lambda_1^s(i-1),\lambda_2^s(i-1))}
% \end{equation}
%where its natural parameters are updated as
\begin{align}
    \lambda_1^{-s}(i)&=\lambda_1(i-1)-\lambda_1^s(i-1),   \\ 
    \lambda_2^{-s}(i)&=\lambda_2(i-1)-\lambda_2^s(i-1).
\end{align}
Next, we compute the tilted distribution $\tilde{g}^s(X_n)$ in \eqref{eq: tilt} using the fast Rao-Blackwellised Gibbs sampler in Section \ref{sec:Rao-Blackwellised Gibbs},
%. 
%Since it is intractable, we use the fast Rao-Blackwellised Gibbs sampler in Section \ref{sec:Rao-Blackwellised Gibbs} to infer it, and the tilted distribution 
which is a Gaussian mixture 
%in \eqref{eq:tilted approximation}, with a form of a mixture of $N_p$ Gaussian components:
\begin{equation}
\label{eq:tilted_mixture}
\tilde{g}^s(X_{n}) 
\;=\; 
\frac{1}{N_p} \sum_{p=1}^{N_p}
\mathcal{N}\!\Bigl(X_{n} \;\Big|\; \tilde{\mu}^{(p)},\, \tilde{\Sigma}^{(p)}\Bigr).
\end{equation}
Then, the first two moments of $\tilde{g}^s(X_n)$ is computed by
%Then, the mean and covariance of $g^{new}(X_n)=\mathcal{N}(X_{n};\mu^{new}(i),\Sigma^{new}(i))$ is set to match the first two moments of $\tilde{g}_s(X_n)$, that is, $\mu^{new}(i)=\tilde{\mu}(i), \ \Sigma^{new}(i)=\tilde{\Sigma}(i)$, and 
% \begin{equation}
%     \mu^{new}(i)=\tilde{\mu}(i), \ \Sigma^{new}(i)=\tilde{\Sigma}(i) ,
% \end{equation}
%where the first and second moments can be computed as
%from the Gaussian mixture in \eqref{eq:tilted_mixture}:
\vspace{-0.5em}
\begin{equation}
\label{eq:tilde_mean}
\tilde{\mu} (i)
\;=\; 
\frac{1}{N_p} \sum_{p=1}^{N_p} \tilde{\mu}^{(p)},
\end{equation}
\vspace{-0.9em}
\begin{equation}
\label{eq:tilde_cov}
\tilde{\Sigma} (i)
\;=\;
\frac{1}{N_p}\,\sum_{p=1}^{N_p}
\Bigl[
  \tilde{\Sigma}^{(p)}
  \;+\;
  \bigl(\tilde{\mu}^{(p)} - \tilde{\mu}\bigr)\,
  \bigl(\tilde{\mu}^{(p)} - \tilde{\mu}\bigr)^\top
\Bigr].
\end{equation}
For computational tractability, we introduce an approximation that assumes covariance matrix for each target $k=1,...,K$ is independent. Thus, $\tilde{\Sigma} (i)$ can be computed as follows:
%, i.e.,  equation \eqref{eq:tilde_cov} can be computed as following for each object $k=1,...,K$:
\begin{equation}\notag
%\label{eq:tilde_cov_eachobject}
\tilde{\Sigma}^k (i)
\;=\;
\frac{1}{N_p}\,\sum_{p=1}^{N_p}
\Bigl[
  \tilde{\Sigma}^{(p),k}
  \;+\;
  \bigl(\tilde{\mu}^{(p),k} - \tilde{\mu}^k\bigr)\,
  \bigl(\tilde{\mu}^{(p),k} - \tilde{\mu}^k\bigr)^\top
\Bigr]. \\[-0.5em]
\end{equation}
Then, the mean and covariance of $g^{new}(X_n)=\mathcal{N}(X_{n};\mu^{new}(i),\Sigma^{new}(i))$ is set to match the first two moments of $\tilde{g}^s(X_n)$. %, that is, $\mu^{new}(i)=\tilde{\mu}(i), \ \Sigma^{new}(i)=\tilde{\Sigma}(i)$. 
By rewriting it into exponential family form $g^{new}(X_n;\lambda_1^{new}(i),\lambda_2^{new}(i))$, 
the site approximation  $g_s(X_n; \lambda_1^s(i),\lambda_2^s(i))$ can be updated by using \eqref{eq:updated site approximation}.
% \begin{equation}
%        g_s(X_n; \lambda_1^s(i),\lambda_2^s(i)) \propto \frac{g^{new}(X_n;\lambda_1^{new}(i),\lambda_2^{new}(i))}{g_{-s}(X_n; \lambda_1^{-s}(i),\lambda_2^{-s}(i))}. 
% \end{equation}
Finally, the natural parameter of the global approximation $ g(X_n; \lambda_1(i),\lambda_2(i))$ can be updated by using \eqref{eq:global apprx}.

Note that Step 2 in Algorithm \ref{alg:EP} is case-dependent: if the network is fully connected, we can directly pass natural parameters of  $g_s(X_n; \lambda_1^s(i),\lambda_2^s(i))$ to all other sensors. When the overall sensor network structure is unknown or communication is limited to neighbouring sensors, various consensus schemes, such as the flooding algorithm \cite{li2016convergence}, can be employed to broadcast local parameters across the network and thus achieve a fully decentralised implementation. To further lower  communication frequency, we design a  variant that performs flooding only once per EP iteration to neighbouring sensors, in which each sensor broadcasts both its current site update and site update from neighbouring sensors from the previous iteration. Over successive EP iterations, each node will progressively receive the recent site updates from all other sensors. Evaluation of this variant will be given in results.

\section{Results}
This section presents the performance evaluation of the proposed distributed EP (DEP) algorithms for multi-sensor tracking. The DEP is evaluated under the assumption that each sensor can communicate with all others, either through a central node or a fully connected network. We also evaluate a variant, distributed EP with one-time flooding (DEP-F). Comparisons are also made with the centralised Rao-Blackwellised Gibbs sampling tracker (C-Gibbs) from \cite{li2023adaptive} and the decentralised gradient variational multi-object tracker with gradient tracking (DeNG-VT-GT) from \cite{li2024decentralised}.

Here we evaluate two simulated datasets, each having 50 Monte Carlo runs with different measurement and trajectory sets. In dataset 1, the network consists of 5 sensors with fixed connectivity% as shown in Fig. \ref{fig:simu1},
, all observing 5 targets in the area. The target Poisson rate for sensor $s$ equal $2s$, $s=1,...,N_s$, and the clutter rate are set to $500$ for all sensors. Dataset 2 presents a more challenging scenario with dynamic network connectivities and increased clutter levels.  The target and clutter rates are set to 2 and 1000, respectively, for all sensors. For all datasets, the total time steps are 50, and the time interval is $\tau=1$s. Objects follow a constant velocity model, with 
${F}_{n,k}^{d} = \begin{bmatrix} 1 & \tau \\ 0 & 1 \end{bmatrix}$,
${Q}_{n,k}^{d} = 36 \begin{bmatrix} \tau^3 / 3 & \tau^2 / 2 \\ \tau^2 / 2 & \tau \end{bmatrix}, d = 1,2$, and $R_k^s=100\text{I}$ where I is a $2$-D identity matrix.  One example sensor network, measurement set and true track of dataset 2 are shown in Figure 1.%\ref{fig:simu1}. 

We use the GOSPA metric \cite{rahmathullah2017generalized} to evaluate the tracking accuracy, where the order $p=1$, $\alpha=2$, and the cut-off distance  $c=50$. A mean GOSPA is computed in Table \ref{tab:performance_compared_methods} that averaged over all sensors, time steps and Monte Carlo runs. Additionally, we use the communication iteration (CI) metric%for comparing the communication cost, 
, which is the total iteration number that sensors pass messages to its neighbours at a time step. DEP and DEP-F use 60 samples with 10 samples burn-in time. %for the Rao-Blackwellised Gibbs sampler.
C-Gibbs method is evaluated with two configurations with subscript denoting the sample size. %: 50 samples with 10 burn-in samples, and 200 samples with 50 burn-in samples. 
%The GOSPA metric also returns submetrics including the localisation errors for well-tracked objects, and the missed/false object errors for disappearance or appearance of objects here it is used for describing the track loss under the setting of a fixed number of objects. 
From Table \ref{tab:performance_compared_methods}, we can see that %at a simpler scenarios 
in dataset 1, all methods have equivalent tracking accuracy, with the proposed EP methods %having the advantage of 
having a lower CI. In a more difficult case of dataset 2, C-Gibbs with a large enough sample size achieves the highest tracking accuracy, but it operates in a centralised setting. In comparison, DEP requires fewer iterations with a generally higher accuracy than DEP-F, though it relies on the fully connected network; otherwise, DEP will need to wait for flooding consensus at each EP iteration. %, making it unsuitable for distributed applications. 
Among distributed methods, DEP-F exhibit lower communication costs than DeNG-VT-GT, though the latter achieves slightly better accuracy. It thus demonstrates that DEP-based methods provide a good trade-off between communication efficiency and tracking accuracy, which is particularly useful in scenarios with constrained communication frequency.

\begin{figure}[tp!]
\centerline{\includegraphics[width=6cm]{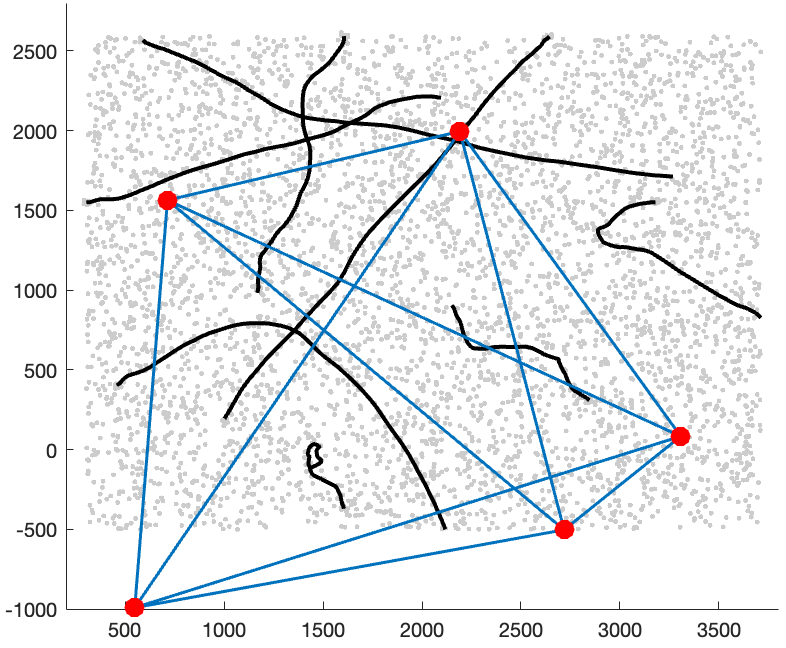}}\label{fig:simu1}
\caption{Example network, measurements and true tracks; grey dots are total measurements at one time step, black lines are true tracks of 8 objects; red dots are sensors and lines denote connectivity at a single time step. }
\vspace{-1.2em}
\end{figure}

\begin{table}[tp!]
\centering
\caption{Performance of compared methods in dataset 1 and dataset 2}
\setlength{\tabcolsep}{3pt} % Adjust space between columns
\begin{tabular}{*5c}
\toprule
Method & \multicolumn{2}{c}{Dataset 1} & \multicolumn{2}{c}{Dataset 2} \\
\cmidrule(lr){2-3} \cmidrule(lr){4-5}
& Mean GOSPA & CI & Mean GOSPA  & CI \\
\midrule
C-Gibbs$_{50}$  & 11.5 $\pm$ 0.4 & -- &36.0 $\pm$ 7.0 & -- \\
C-Gibbs$_{200}$  & 11.5 $\pm$ 0.4 & -- & 33.8 $\pm$ 2.0 & -- \\
DEP  & 11.5 $\pm$ 0.4 & 5 & 35.0 $\pm$ 5.2 & 5 \\
DEP-F  & 11.5 $\pm$ 0.4 & 5 & 36.7 $\pm$ 8.3 & 10 \\
DeNG-VT-GT  & 11.5 $\pm$ 0.4 & 50 & 34.5 $\pm$ 4.2 & 50 \\
\bottomrule 
\end{tabular}
\label{tab:performance_compared_methods}
\vspace{-1.2em}
\end{table}

% \begin{table}[htp!]
% \centering
% \caption{Performance of compared methods in dataset 2}
% \setlength{\tabcolsep}{3pt} % Adjust space bet columns
% \begin{tabular}{*3c}
% \toprule
% method & Mean GOSPA   & communication iteration \\
% \midrule
% C-Gibbs$_{50}$  & 35.97 $\pm$ 6.95 & --   \\
% C-Gibbs$_{200}$  & 33.77 $\pm$ 1.94 & --   \\
% %C-VT  & 35.18 $\pm$ 7.89 & --   \\
% DEP$_{50}$  & 35.03 $\pm$ 5.15 & 5   \\
% DEP$_{200}$  &  $\pm$  & 5   \\
% DEP-F$_{50}$  & 37.22 $\pm$ 8.29 & 5   \\
% DEP-F$_{50}$  & 36.66 $\pm$ 8.31 & 10   \\
% DEP-F$_{200}$  &  $\pm$  & 10   \\
% DeNG-VT-GT  & 34.53 $\pm$ 4.17 & 50   \\
% \bottomrule 
% \end{tabular}
% \label{Performance of compared methods}
% \end{table}

\section{Conclusion}
This paper introduces, for the first time, a novel distributed EP algorithm for multi-sensor, multi-object tracking in cluttered environments. The proposed EP efficiently approximates the posterior distribution of shared target states while marginalising out local sensor-specific data associations. By enabling each sensor to operate independently and exchange only moment estimates, the approach significantly reduces computational complexity and enhances scalability in distributed networks. Future research will extend the framework to accommodate an unknown and time-varying number of objects.

\bibliographystyle{IEEEtran}
\bibliography{./IEEEabrv,IEEEexample}

\end{document}